\documentclass[aip,amsmath,amssymb,reprint, twocolumn]{revtex4-1}
\usepackage[utf8]{inputenc}
\usepackage[T1]{fontenc}
\usepackage{mathptmx}
\usepackage{listings}
\usepackage{xcolor}
\usepackage{siunitx}
\usepackage{lipsum}
\usepackage{wrapfig}
 
\lstdefinestyle{mystyle}{
    backgroundcolor=\color{backcolour},   
    commentstyle=\color{codegreen},
    keywordstyle=\color{magenta},
    numberstyle=\tiny\color{codegray},
    stringstyle=\color{codepurple},
    basicstyle=\ttfamily\footnotesize,
    breakatwhitespace=false,         
    breaklines=true,                 
    captionpos=b,                    
    keepspaces=tru
    numbersep=5pt,                  
    showspaces=false,                
    showstringspaces=false,
    showtabs=false,                  
    tabsize=2
}
 
\usepackage{times,amsmath}
\usepackage{epsfig}
\usepackage{color}
\usepackage{longtable}
\usepackage{ulem}
\usepackage{graphicx}% Include figure files
\usepackage{dcolumn}% Align table columns on decimal point
\usepackage{bm}% bold math
\usepackage{bookmark}
\usepackage{tabularx}% bold math
\usepackage{hyperref}
\usepackage{multirow}

\usepackage{tocloft}
\hypersetup{colorlinks=true, citecolor=blue, filecolor=blue, linkcolor=blue, urlcolor=blue}

\usepackage[switch, pagewise]{lineno}
\linenumbers\relax % Commence numbering lines
\nolinenumbers

\begin{document}

\title{Quantification of Crystal Packing Similarity from Spherical Harmonic Transform}

\author{Qiang Zhu}
\email{qiang.zhu@unlv.edu}
\affiliation{Department of Physics and Astronomy, University of Nevada Las Vegas, NV 89154, USA}

\author{Weilun Tang}
\affiliation{Department of Physics and Astronomy, University of Nevada Las Vegas, NV 89154, USA}

\author{Shinnosuke Hattori}
\affiliation{Advanced Research Laboratory, R\&D Center, Sony Group Corporation, 4-14-1 Asahi-cho, Atsugi-shi 243-0014, Japan}

\date{\today}
\begin{abstract}
In this work, we present a new computational approach to characterize and classify molecular packing in the solid states. The key idea is to project each neighboring molecule (or short contact) from the centered molecule into a unit sphere according to the interaction energy. Consequently, the similarity between two spherical images can be evaluated from the spherical harmonics expansion based on the maximum cross-correlation. We apply this approach to successfully reproduce the previous packing assignment on a small amount of data with an improved categorization. Furthermore, we conduct a packing similarity analysis over 2000 hydrocarbon crystal data sets and uncover a set of abundant packing motifs. Unlike the previous approaches based on the subjective visual comparison at the real space, our approach provides a more robust way to measure the packing similarity, thus paving the way for a rapid classification of large scale crystal data.
\end{abstract}

%\noindent\textbf{Keywords:} crystalline, materials genome, modeling, van der Waals, phase transformation\\\\

\maketitle
\makeatletter

%\lipsum
%\twocolumn
%\tableofcontents
%\listoffigures
%\listoftables
%\linenumbers
\section{INTRODUCTION}

Molecular solids have been extensively used as key components in chemical industries such as medicine \cite{lee2011crystal}, fertilizers, dyes \cite{zhuo2022organic}, pesticides \cite{Yang-ANIE-2017}, and high energy explosives \cite{liu2018polymorphism}, as well as electronics industry since the discovery of bulk conductivity in polycyclic aromatic compounds in 1950s \cite{kallmann1960bulk}. While some molecules can aggregate with no particular order, such as amorphous solids, most organic solids are crystalline, and their physical properties largely depend on regularly repeating intermolecular packing. Nevertheless, understanding the packing of molecules is a rather elusive subject due to the variety of molecular shapes. Unlike the packing of spheres in the atomic crystals, geometry analysis on the irregularly shaped molecules appears to be more of an art than a science. It is visually difficult to capture the pattern even for experienced researchers \cite{day2006experiment}. 

\begin{figure}[ht]
\centering
\includegraphics[width=0.48 \textwidth]{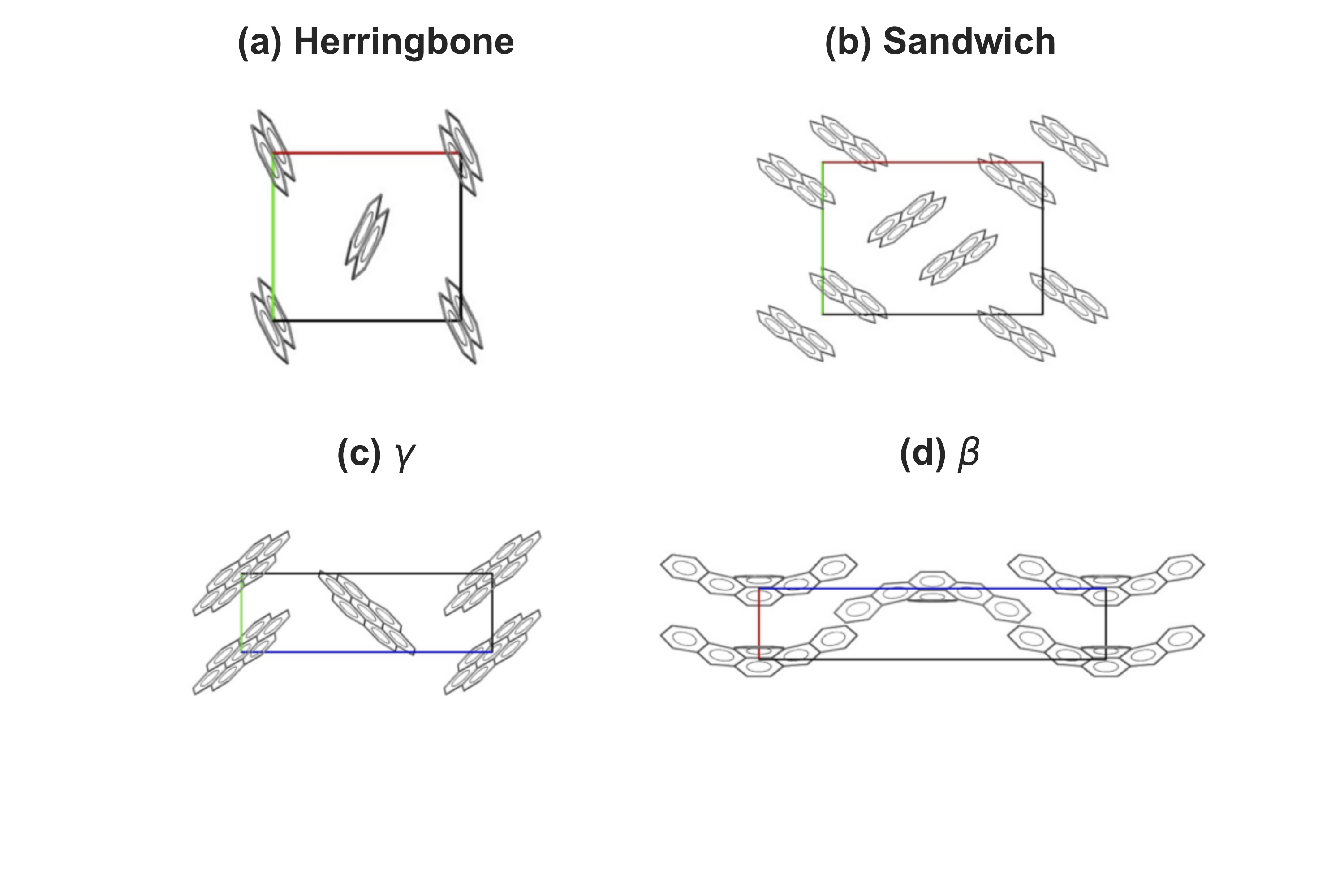}
\caption{\label{Fig1} Four packing motifs on the crystals made of planar aromatic hydrocarbons (PAHs).}
\end{figure}

To ease the visual challenge, most early works focused on the simple shapes. Among them, Robertson first proposed to divide planar aromatic hydrocarbons (PAH) into two categories based on the ratio of molecular area to the thickness \cite{Robertson-1951}. It was found that the disk-like molecules tend to stack together via rigid translation (\textit{stack-promoting}), while the molecules with smaller areas prefer the glide-like reflections (\textit{glide-promoting}). Using energetic as well as geometrical criteria, Desiraju and Gavezotti \cite{desiraju1989crystal} extended the categorization of PAHs into four groups and used them as a guide for prediction of crystal packing for new molecules. As shown in Fig. \ref{Fig1}, they include (i)\textit{herringbone}, in which each molecular column has interactions with the nonparallel neighboring molecules; (ii) \textit{sandwich}, consisting of the herringbone motif with sandwich-type diads; (iii) $\gamma$, the flattened-out herringbone; and (iv) $\beta$, a layered structure made up of graphite-like planes.

The packing motif concept has been applied to establish the correlations between these materials’ packing and observed physical properties (e.g., charge transport for organic semiconductors \cite{Campbell-JMC-2017, zhang2018theoretical} and insensitivity for molecular explosives \cite{mathieu2017sensitivity}). Despite its popularity in crystal analysis, the definition of four patterns, like many other chemical nomenclatures, lacks a mathematical rigor. From the definition, it is hard to define the boundary between $\gamma$ and $\beta$. Even within the herringbone group, some crystals clearly have different packing as compared to others. Although several tools have also been developed to automate the assignment of packing motif \cite{Campbell-JMC-2017, loveland2020automated, ito2020estimation}, they sometimes yield inconsistent results due to the subjective choice of projection plane or simply due to the ambiguity of the definition itself. 

In addition, the Hirshfield fingerprint has been commonly used to analyze the complex information contained in a molecular crystal structure into a single, unique full colour plot \cite{Mckinnon:2007}. Derived from the Hirshfeld surface, these 2D-fingerprint plots provide a visual summary of the frequency of each combination of internal and external distances across the surface of a molecule, so they do not only indicate which intermolecular interactions are present, but also the relative area of the surface corresponding to each kind of interaction. Other analysis techniques \cite{motherwell2010molecular, Spackman-2016, carugo2017packing}  were also proposed recently. However, the interpretation is either too abstract or too cumbersome, and they also lack a simple criterion to judge if two crystals are truly similar or not. On the other hand, the COMPACK algorithm \cite{compack} is commonly used to check if two structures are identical or not when they consist of the similar molecules. Nevertheless, it is not suitable to compare the similarity between two crystals with different constituents, which is important in many crystal engineering applications.  

To address the aforementioned challenge, we introduce the image approach in combination with spherical harmonics expansion to quantify the packing similarity of organic crystals. This essentially involves a two-step process. First, we construct the spherical images to represent the molecular packing and energy distribution in a crystal. Second, the similarity of two spherical images (i.e., crystal packing) is computed by the maximum normalized cross correlation in the Fourier space. In the following, we present the methodology and its application to a set of hydrocarbon crystal data. Due to its mathematical rigor, this approach can be fully automated for a rapid classification of distinct crystal packing motifs in a large data set,  regardless of the molecular choices. 

\section{Computational Methodology}

\subsection{Image Representation of Molecular Packing}
\begin{figure}[ht]
\centering
\includegraphics[width=0.48 \textwidth]{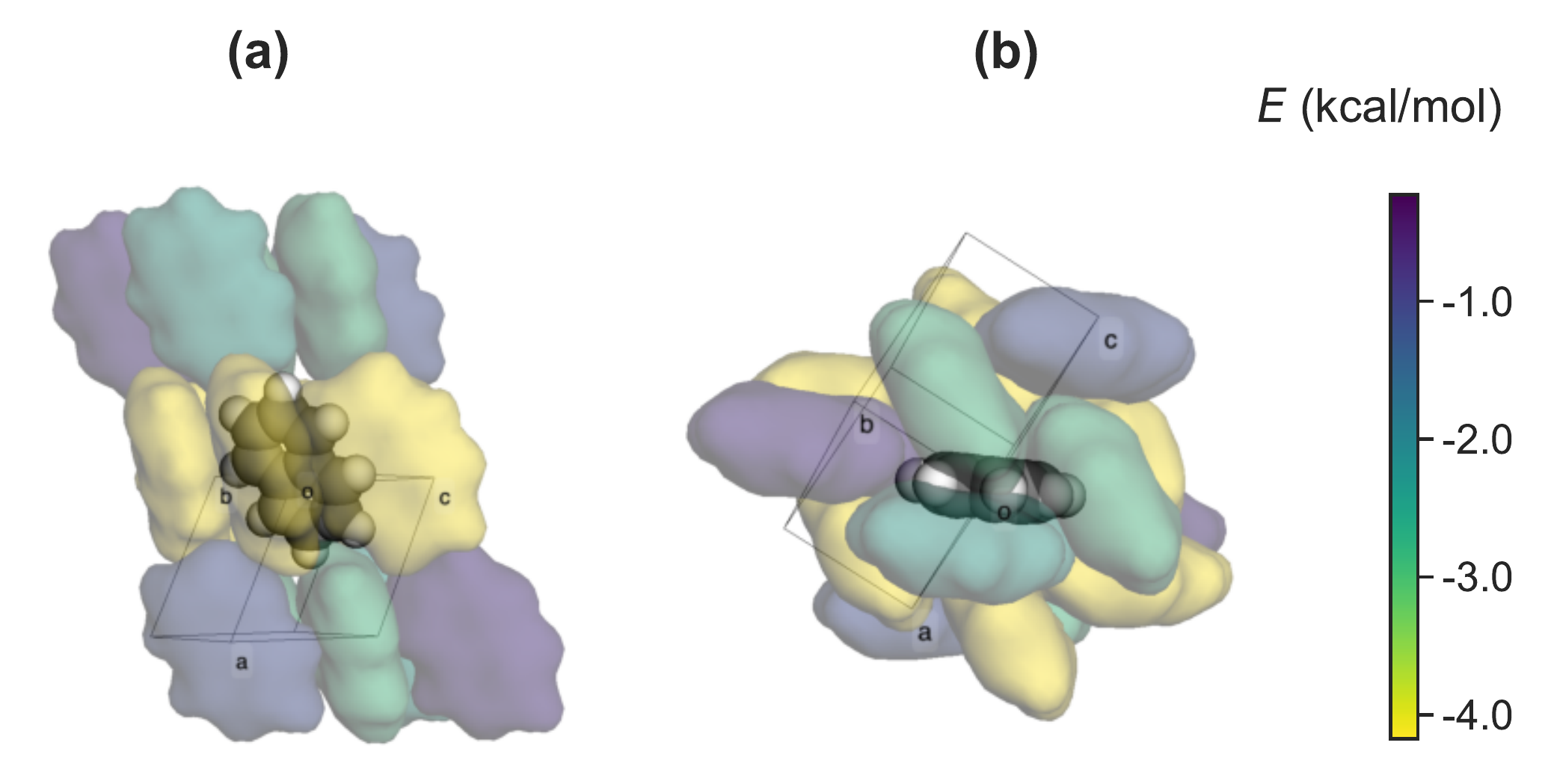}
\caption{\label{Fig2} Different projections of NAPHTA crystal from the top (a) and side (b) views. The center molecule is shown as spheres while the surrounding molecules are shown by their van de Waals surface colored by the interaction energy.}
\end{figure}

Although it is highly subjective to describe the crystal packing, there is a general consensus that at least two factors are of critical importance. First, the geometry has been widely used to understand the packing. The simplest analysis is to count the coordination number as a function of the cutoff distance. Moreover, the spatial distribution of neighboring molecules can also analyzed with more sophisticated tools \cite{BO-PRB-1983, CNA-JPC-1987}. However, crystal packing is not simply a geometry problem. The intermolecular energy should be taken into account as well, as described by Desiraju and Gavezotti \cite{desiraju1989crystal}, as well as the 2D-fingerprint of the Hirshfeld surface \cite{Mckinnon:2007}. Using the naphthalene crystal (referred as NAPHTA in the Cambridge Crystal Database) as an example, a molecule can be generally considered to have 16 neighbors (see Fig. \ref{Fig2}). Based on the intermolecular interactions, these molecules can divided to four tiers, colored in yellow (6), green (6), light blue (2) and deep blue (2). The last tier is often omitted because their interactions are much weaker. One can consider that the most important yellow tier molecules form the equator while the less important green and blue molecules are distributed on the Southern and Northern Hemispheres by following some pattern. Therefore, a good representation should be able to probe such characteristics of the spatial distribution of both molecules and energetics.

Inspired by those previously developed approaches, we construct the spherical image to represent the packing as follows,

\begin{enumerate}
    \item Choose the center molecule ($M_0$) and find the neighboring molecules or intermolecular short contacts $\{M_i\}$.
    \item For each $M_i$, compute the bonding energies ($\{E_i\}$) with respect to $M_0$. If $M_i$ denotes a molecule, $E_i$ is the sum of all interactions between $M_i$ and $M_0$.
    \item Project each $M_i$ onto the unit sphere centered at $M_0$ with the spherical coordinates $d_i = (\theta_i, \phi_i)$. If $M_i$ denotes a short contact (e.g., C$\cdots$C or C$\cdots$H pairs), the center of the atomic pair is used as the reference point to determine $d_i$. When $M_i$ denotes a molecule, the molecular center shall be used as the reference. 
    \item For each $M_i$, place a Gaussian based on $E_i\exp\frac{(d-d_i)^2}{2\sigma^2}$ in the unit sphere grids.
\end{enumerate}

Therefore, we come up with two definitions of spherical functions as follows:
\begin{equation}
    f = \begin{cases}
        \sum\limits_{i=1}^{\textrm{molecules}} E_i \exp \frac{(d-d_i)^2}{2\sigma^2} & \textrm{Coarse grained model}\\
        \sum\limits_{i=1}^{\textrm{contacts}}  E_i \exp \frac{(d-d_i)^2}{2\sigma^2} & \textrm{Fine resolution model}
    \end{cases}
\end{equation}

\begin{figure}[ht]
\centering
\includegraphics[width=0.48 \textwidth]{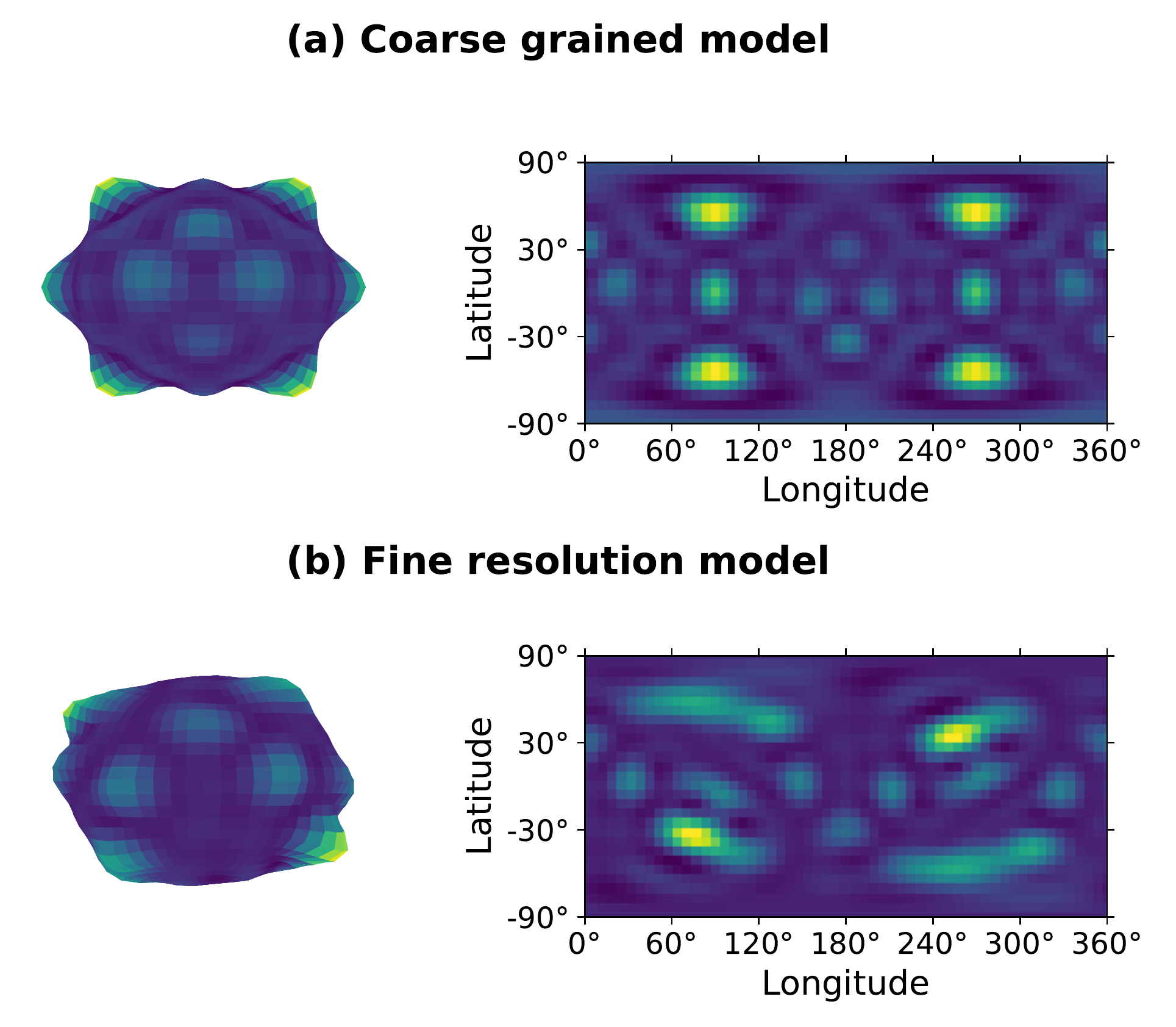}
\caption{\label{Fig3} Two sphere image representations for NAPHTA, (a) the coarse grained model and (b) the fine resolution model.}
\end{figure}

In general, $f_\textrm{molecule}$ can be considered as a coarse-grained descriptor for the molecular packing. Fig. \ref{Fig3}a shows the representation for a naphthalene crystal (see Fig. \ref{Fig2}). Unlike other descriptors such as Hirshfield fingerprint plots \cite{Mckinnon:2007}, the spherical image can be easily interpreted since it retains the spatial information. For instance, the center of each hot spot denotes the position of the neighboring molecule. In NAPHTA, one can clearly find six bright spots at the equator plane, and four less bright spots at each of the hemispheres. This information is consistent with our analysis from Fig. \ref{Fig2}. It is essentially the arrangement of hot spots in the sphere. For such a spherical image, it is also common to plot its cylindrical projection as used in the map of the Earth. On the other hand, $f_\textrm{contact}$, shown in Fig. \ref{Fig3}b is a break-down version of $f_\textrm{molecule}$ or a projection of the Hirshfield surface into the unit sphere \cite{Spackman-2016}. It can capture more details when describing the molecular packing. As one can see in the projection map in Fig. \ref{Fig3}b, some bright spots becomes rather diffusive, indicating that the intermolecular interaction are from multiple short contacts between the center and surrounding molecules, while the very hot spots suggest that there exist very strong interactions such as C$\cdots$H bonds for the PAH system, or hydrogen bonds for the general cases. Therefore, $f_\textrm{contact}$ may be more advantageous for a direct comparison of two different images.  

When computing the intermolecular energy, one can use different approaches from empirical force fields to more advanced electronic structure methods. In our case, we follow an empirical atom-atom potential \cite{gavezzotti1994crystal} due to its simplicity. In addition, it is important to note that the choice of Gaussian smearing is to ensure a smooth distribution when comparing two distinct local environments. As such, it can be also used to characterize the packing environments in an aggregate, amorphous solid, or even liquid. In the following, we will keep $\sigma$ = 0.1 \AA~  at the unit sphere for all the subsequent calculations. To construct the spherical image, we uniformly sampled 10,000 grids on the unit sphere according to the Fibonacci method, which is more efficient than the evenly spaced grid sampling on ($\theta, \phi$).

\subsection{Real Spherical Harmonic Expansion}
For a real valued spherical function $f(\theta, \phi)$, it can be expressed as a series of spherical harmonic functions, 

\begin{equation}
    f(\theta, \phi) = \sum_{l=0}^\infty \sum_{m=-l}^l f_{lm}Y_{lm} (\theta, \phi),
\end{equation}

where $f_{lm}$ is the spherical harmonic coefficient, $Y_{lm}$ is the spherical harmonic function, and $l, m$ are the spherical harmonic degree and order. The details of $Y_{lm}$ are given in the appendix.

Similar to the Fourier transform of a one-dimension spectrum, the purpose of spherical harmonic expansion is to efficiently extract the features from the frequency domain with only a few $f_{lm}$ coefficients. It is straightforward to show that $f_{lm}$ can be calculated by the integral
\begin{equation}
\label{isph}
    f_{lm} = \frac{1}{4\pi} \int _\Omega f(\theta, \phi) Y_{lm} (\theta, \phi) d\Omega
\end{equation}

\begin{figure}[ht]
\centering
\includegraphics[width=0.50 \textwidth]{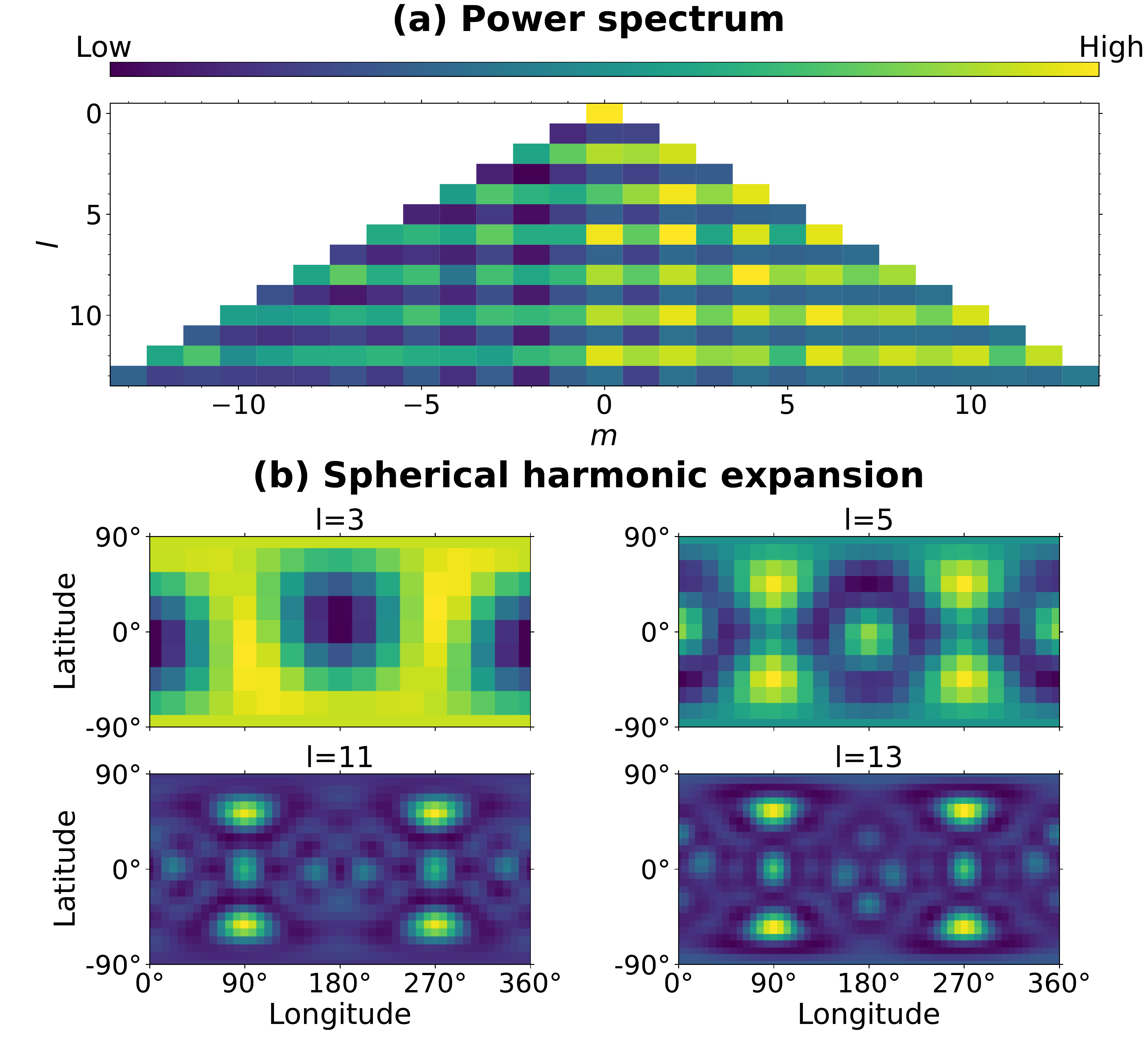}
\caption{\label{Fig4} Spherical harmonic expansion. (a) shows the spectrum as a function of spherical harmonic degree ($l$) and order ($m$. (b) plots the reconstructed images from different cutoff values of $l$.}
\end{figure}

In general, $f_{lm}$ consists of $2\times l\times l$ real valued numbers. Fig. \ref{Fig4}a shows the power spectrum of $f_{lm}$ obtained from the naphthalene crystal. Compared to the grids in the real space, these $f_{lm}$ coefficients provide an economical way to store all information to reconstruct the images. With the increase of spherical harmonic degree $l$, it is expected that more details of the function can be described. However, the $f_{lm}$ at very high order simply means the noise. Therefore, there should be a cutoff when it is found that further increase of $l$ no longer brings any notable improvement.

In practice, we construct the spherical image by sampling 10,000 grids on the unit sphere according to the Fibonacci method, and compute the expansion coefficients from eq. \ref{isph}. Fig. \ref{Fig4}b plots the recovered images from different $l_\textrm{max}$ values. In this work, we choose $l_\textrm{max}$ = 13 that should be sufficient to recover all spherical images in high accuracy.

\subsection{Cross-correlation on the Sphere}
Having established the compact description for one crystal packing ($f$), we can apply the same scheme to describe another crystal named $g$. Obviously, we arrive at the point to address the original question, namely, how to quantify the similarity between $f$ and $g$? It is clear that the use of expansion coefficients is advantageous to reconstruct the original function. Therefore, we wish to compute the similarity based on these coefficients. 

One possible solution is to derive a set of rotation-invariant arrays. For instance, the power spectrum has been popularly used in distinguishing the local environments of an atomic crystal \cite{BO-PRB-1983, Bartok-PRB-2013}. However, it can lead to a substantial dimensionality reduction from the original coefficients ($2\times l\times l$) to a power spectrum ($l+1$), This indicates that a lot of information is lost during this process. Indeed, we found this is not sufficient since many different packing patterns may share a similar power spectrum. 

To avoid unnecessary information loss, we aim to seek a better metric by considering all $f_{lm}$ coefficients when comparing the similarity. For image similarity analysis, a common way is to analyze the cross-correlation spectrum spectrum between two functions $f$ and $g$,
\begin{equation}
    C_{l} (f, g) = \int f(\Omega)g(\Omega)
\end{equation}

From the cross-correlation, we can further define a scalar metric at the Fourier space that is bounded between 0 and 1,
\begin{equation}
    S(f, g) = \frac{\sum\limits_{lm} f_{lm}g_{lm}}
    {\sqrt{\sum\limits_{lm} f^2_{lm} \sum\limits_{lm} g^2_{lm}}}
\end{equation}

\begin{figure}[ht]
\centering
\includegraphics[width=0.50 \textwidth]{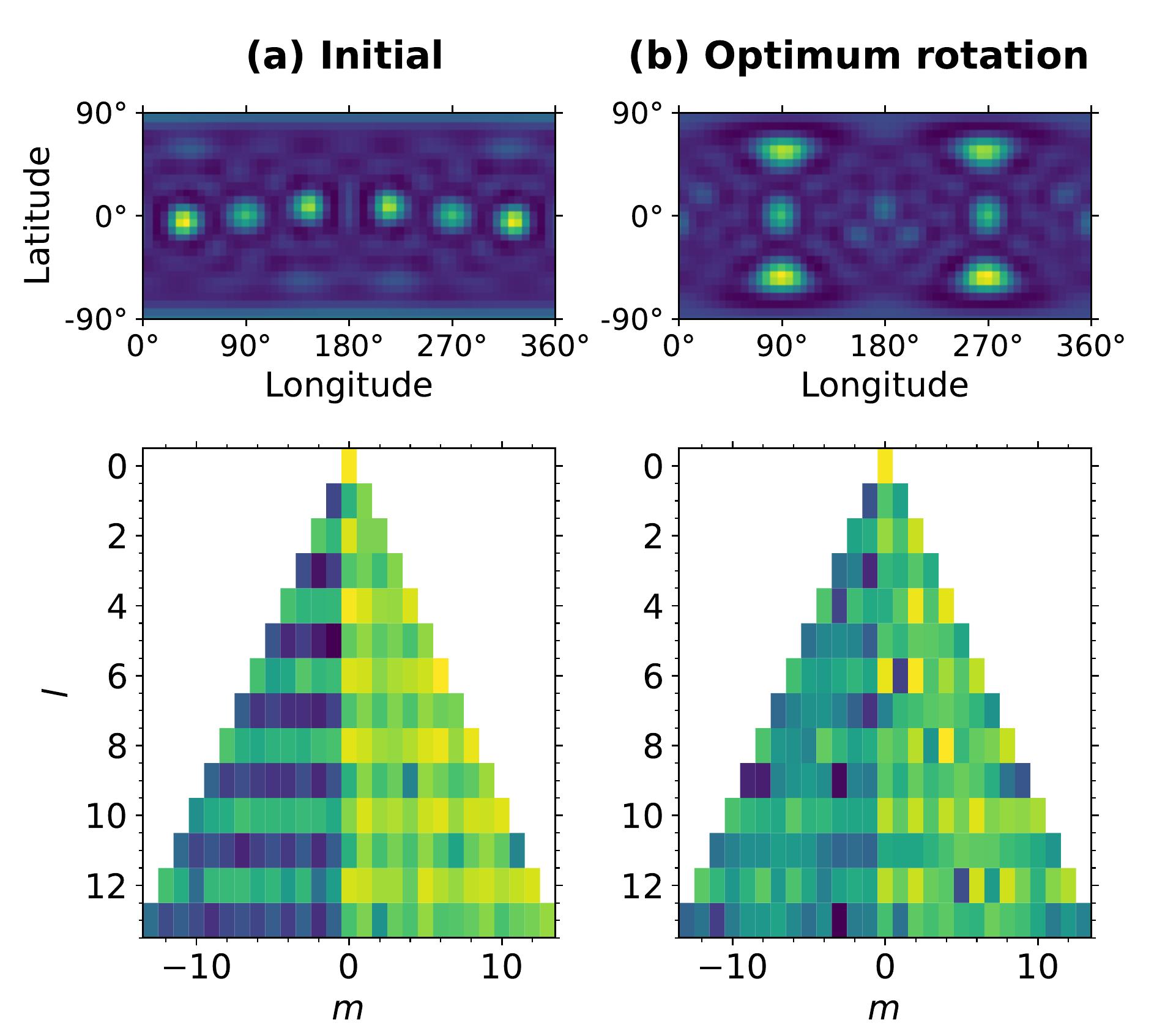}
\caption{\label{Fig5} The spherical function's projection and power spectrum of PHENAN before (a) and after (b) rotation with respect to NAPTHA.}
\end{figure}

According to the definition, $f$ and $g$ is identical if $S=1$. When $S$ has a strong deviation from 1, it means $f$ and $g$ are less similar. However, it is important to note that the coefficients are subject to change under a rotation ($R$, which can also be denoted by a set of Euler angles $\{\alpha, \beta, \gamma\}$). Therefore, we seek to maximize $S$ by sampling all possible rotational space of the SO(3). In practice, we perform deterministic quasi random sampling from the low-discrepancy Sobol sequence \cite{joe2008constructing} to generate a uniform grids of $\{\alpha, \beta, \gamma\}$ on the sphere. Then we rotate $g$ accordingly, and $S$ is further maximized based on the derivative-free optimization method as implemented in the Scipy.optimize tool box \cite{scipy}. Finally, the search returns the $S_{\textrm{max}}(f, g)$, a scalar number between 0 and 1, as well as the rotation ($R$) on $g$ that achieves the best match. Fig. \ref{Fig5} shows the initial and optimized rotation for PHENAN when it is compared with naphthalene by following this procedure. For the initial state in Fig. \ref{Fig5}a, it returns $S=0.253$. However, the true $S$ becomes 0.931 after the best rotation is found.

\section{Results}
With this new image approach, we proceed to build the first link between inorganic and organic crystallography. Considering hydrocarbon molecules as the simplest category, like elemental allotropes in the atomic crystals, can we classify them by some packing motif? If so, what will be the most common motifs, like body-centered-cubic (bcc), face-centered-cubic (fcc), hexagonal-close-packing (hcp) and diamond types? In the following, we will first validate the approach in a small set of previously investigated systems and then present our own categorization for a more extended data consisting of over 2000 crystals.

\subsection{Regroup of 30 PAH crystals}
\begin{figure*}[ht]
\centering
\includegraphics[width=0.9 \textwidth]{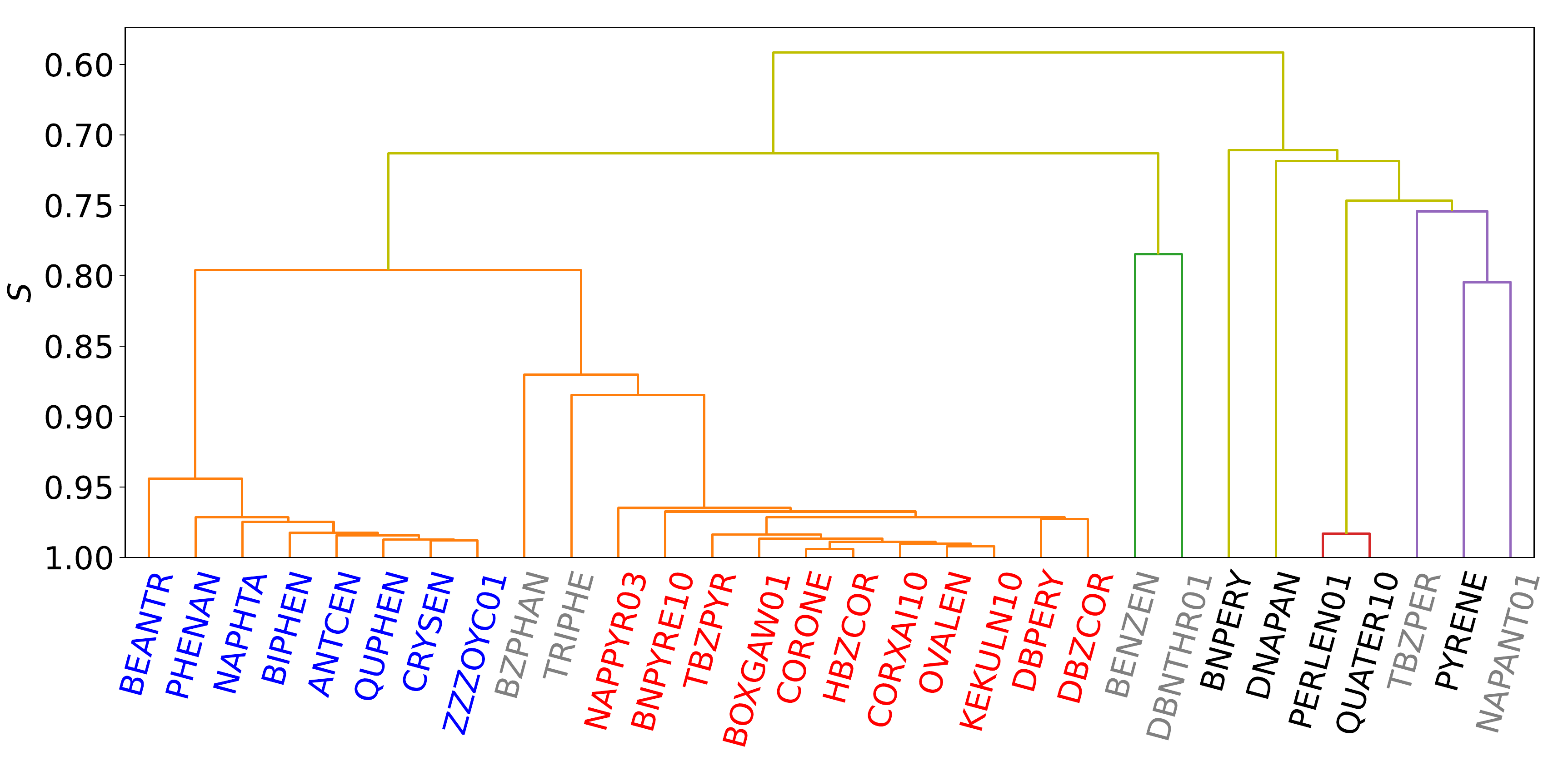}
\caption{\label{Fig6} The hierarchical clustering of 30 PAH crystals. The structures belonging to herringbone, sandwich are colored in blue and black, respectively. The $\gamma$ and $\beta$ types are indistinguishable, and hence they are all colored in red. In addition, several structures that cannot be matched to the previous assignments are colored in grey.}
\end{figure*}

In 1989, Desaraju and Gavezotti have defined four prototypical packing for about 30 crystals of disk-shaped PAH molecules \cite{desiraju1989crystal}. Their pioneering work has significantly influenced the following studies in crystal engineering. With this new image approach, we revisited this data set. We calculated the similarity function for each structure pair. From the computed similarity matrix, we performed an unsupervised hierarchical agglomerative clustering, Fig. \ref{Fig6} displays the resulting dendrogram plot of the 30 PAH crystals. Clearly, our results largely agree with the previous assignment \cite{desiraju1989crystal}. Out of 30 samples, we found that 24 of them are closely clustered into three groups, corresponding to the previously assigned herringbone, sandwich and $\gamma/\beta$ types. This is encouraging, given that our calculation is fully automated without any supervision. From these results, we can also analyze each cluster in detail.

\textbf{Herringbone} is the most common group. Using NAPHTA as the reference, we found that most crystals within this group (colored in red in Fig. \ref{Fig6}) have rather high $S$ values, including CRYSEN (0.975), ANTCEN (0.974), ZZZOYC01 (0.945), QUPHEN (0.941), BIPHEN (0.938), PHENAN (0.931), BEANTR (0.896). 

\textbf{Sandwich} was the least common group with only five examples in its original proposal \cite{desiraju1989crystal}. Compared to the herringbone group, the $S$ values between each member are smaller. However, they can be clearly grouped to the same big branch (colored in black in Fig. \ref{Fig6}) with the unsupervised clustering method, indicating a strong correlation. According to our analysis, there actually exist two types of sandwich crystals. The details will be discussed in the following section.  

$\bm{\gamma/\beta}$ are almost identical if one only focuses on the arrangement of molecular centers in Figs. \ref{Fig1}c-d. They can be only distinguished by counting the ratio of C$\cdots$C and C$\cdots$H interactions \cite{desiraju1989crystal}. Indeed, the converted coarse grained tends to group them together (colored in red in Fig. \ref{Fig6}) with the high threshold value among three major clusters. However, one can always use the fine resolution model to search for really matched crystals.

Despite the overall agreement, some structures, colored in grey in Fig. \ref{Fig6} cannot be assigned to the same groups as suggested in the previous literature. For instance, the benzene crystal (BENZEN) was grouped to the herringbone type, but the authors also mentioned that it was more like an outlier \cite{desiraju1989crystal}. Similarly, several other crystals (including BZPHAN, TRIPHE, DBNTHR01, TBZPER, NAPANT01) clearly exhibit different patterns from any of four types as shown in Fig. \ref{Fig1}. The differences can be directly detected by a trivial visual analysis. Therefore, our image approach is more advantageous since it can provide a robust way to detect such outliers without ambiguity. %Hence it is advantageous over the previous methods. 

\subsection{Common prototypes on a larger data set}
Encouraged by the successful application on the 30 PAH crystals, we systematically inquired all crystals from the Cambridge Structural Database by searching for the systems containing only C and H elements and no more than one molecule in the asymmetric unit. After removing the duplicate entries, we obtained 2007 crystals (see more details in the supplementary materials). This data set serves as the test bed to further validate our image approach.

\begin{figure*}[ht]
\centering
\includegraphics[width=0.95\textwidth]{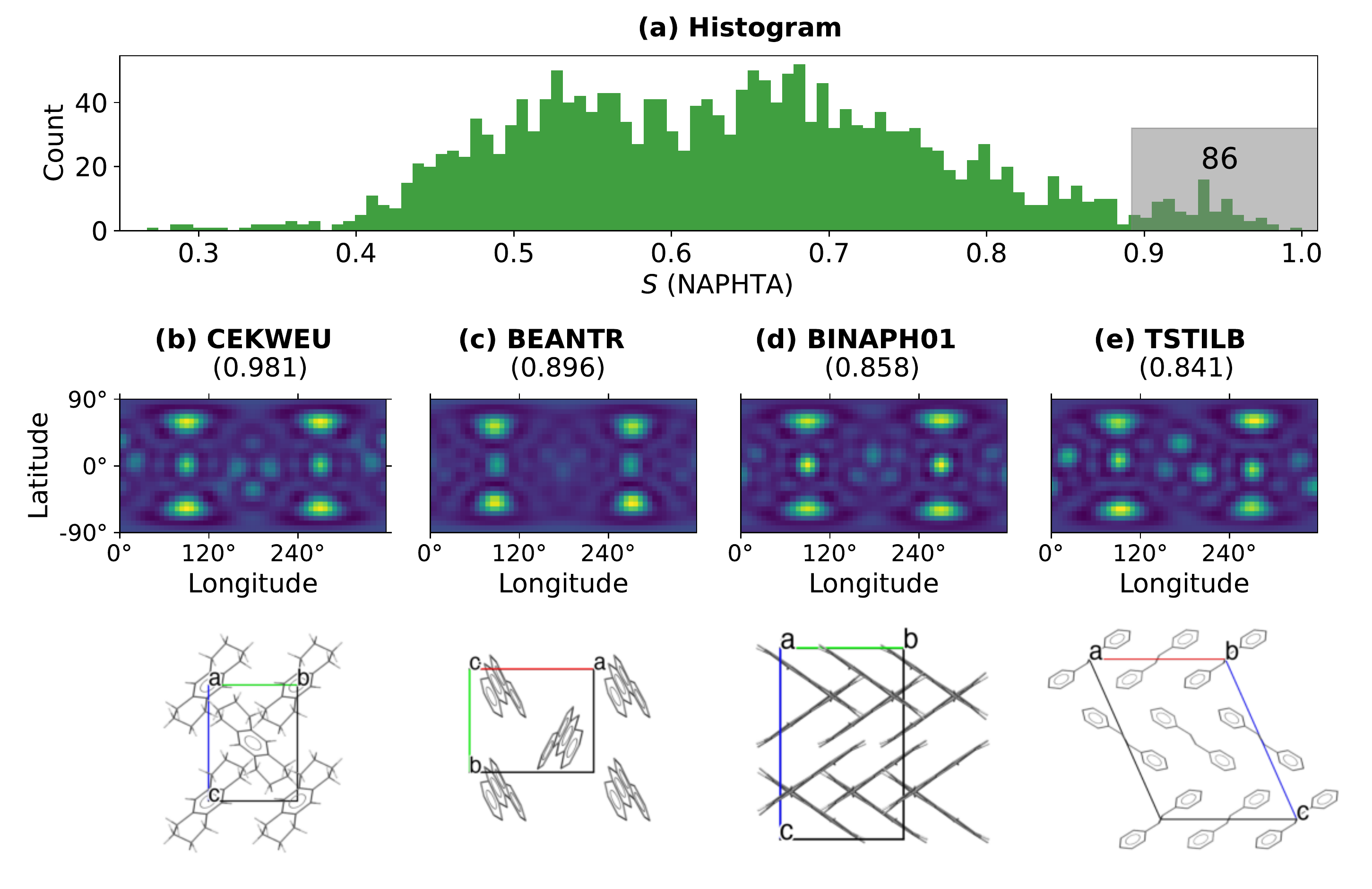}
\caption{\label{Fig7} The naphthalene crystal. (a) is the similarity ($S$) distribution of all PAH crystals with respect to NAPHTA. (b)-(e) show several representative crystals with different $S$ values.}
\end{figure*}

\subsubsection{The naphthalene family}
In the early days of small molecule crystallography, the naphthalene crystal was considered to be the common salt of organic crystal \cite{kitaigorodsky2012molecular}. This type of crystal packing was assigned to the herringbone group, which was found to occur more often than any other group. Fig. \ref{Fig7}a plots the histogram of the distances between NAPHTA with all other hydrocarbon crystals. The entire distribution can be roughly described by some normal distribution with a mean around 0.6-0.7 (more details can be found in the supplementary materials). Here we focus on the region with high $S$ values. Among the entire data set, we found CEKWEU (Fig. \ref{Fig7}b) has the best match with NAPHTA. Although the molecule in CEKWEU is not flat, the mapping is obvious as they share the same space group ($P2_1/c$) and Wykcoff position, as well as similar molecular orientation. In addition, Figs. \ref{Fig7}c-e displays several structures with different $S$ values. By analyzing these patterns as well as the 3D structure, we decide to set a threshold of 0.892, leading to 86 crystals belonging to the naphthalene family. 

It is important to note that our similarity calculation does impose any symmetry constraints. As shown in Table \ref{table1}, several space groups, e.g.,  $P2_1/c$, $Pbca$, $P\overline{1}$, $P2_1$, occur more often. Clearly, all of them follow group-subgroup relations. For instance, the molecules containing four or more aromatic rings crystallize in the $P\overline{1}$ with Z=2 (e.g., TENCEN, PENCEN), but the overall packing is still close to that of NAPHTA. It would be interesting to collect more data and use them to predict the possible phase transition.  

\begin{table}[]
\caption{The space group distribution of 86 naphthalene crystals.}\label{table1}
\begin{tabular}{ccc}
\hline\hline
~~Space group~~     & ~~\# of molecules per cell ~~& ~~Occurrence~~ \\ \hline\hline
$P2_1/c$        & 2                        & 34        \\ 
$Pbca$          & 4                        & 14        \\ 
$P2_1/c$        & 4                        & 9         \\ 
$P\overline{1}$ & 2                        & 7         \\ 
$P2_1$          & 2                        & 5         \\ 
$Pnma$          & 4                        & 5         \\ 
$P2_12_12_1$    & 4                        & 3         \\ 
$Aea2$          & 4                        & 2         \\ 
$C2/c$          & 4                        & 2         \\ 
$Pbcn$          & 4                        & 2         \\ \hline\hline

\end{tabular}
\end{table}

Finally, we emphasize that the coarse-grained image model is still limited by its molecular approximation. Therefore, it is not able to detect the difference in terms of molecular shape. This can be corrected by the fine-resolution model. If we simulate the similarity based on $f_\textrm{contact}$, only three crystals ANTCEN (0.917), CEKWEU (0.834), DMANTR (0.833) can match NAPHTA well. Therefore, we suggest the use of $f_\textrm{contact}$ when it is necessary to find a really good match on a reference crystal.

\begin{figure}[ht]
\centering
\includegraphics[width=0.5\textwidth]{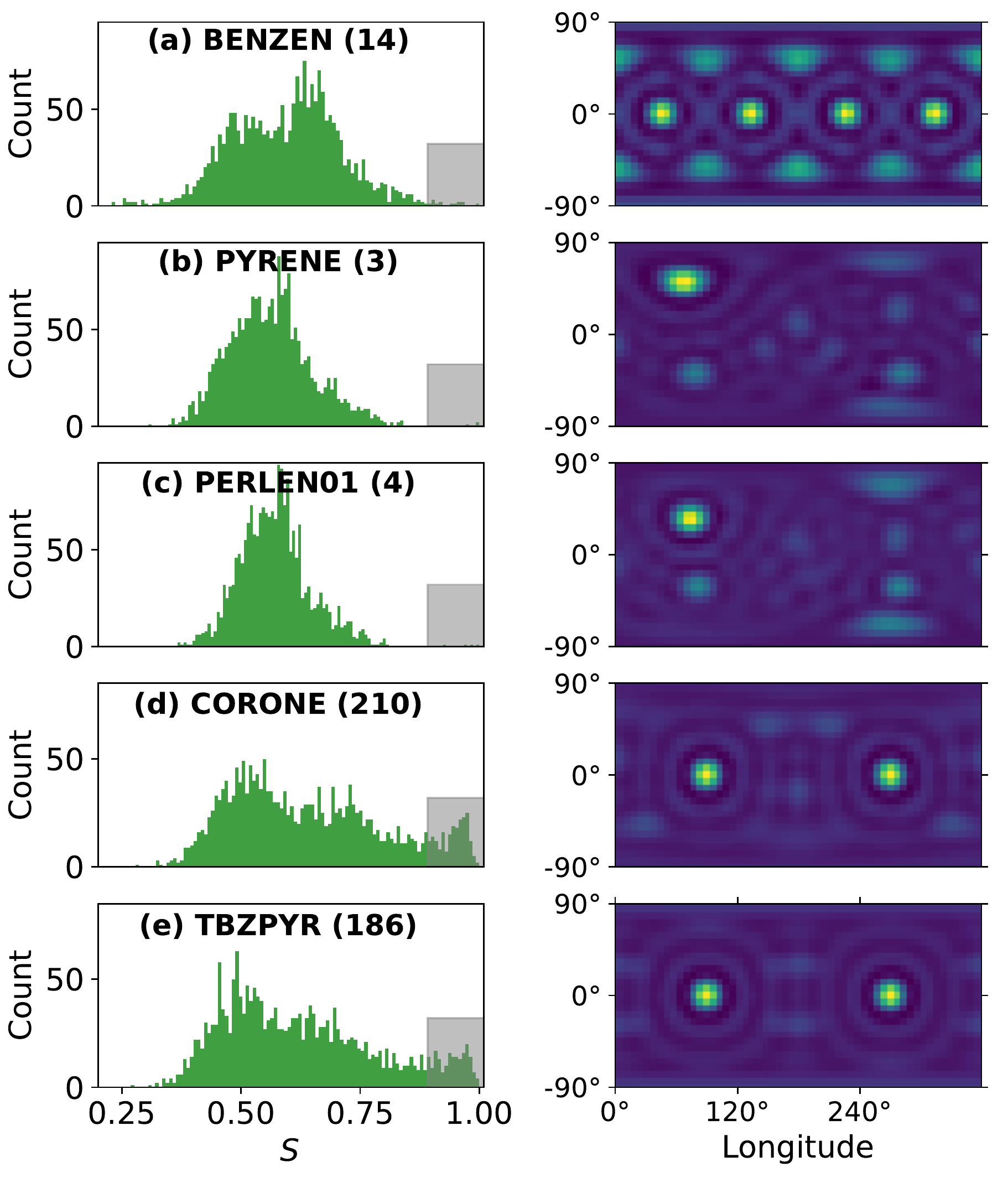}
\caption{\label{Fig8} Other representative crystals, including (a) benzene type, (b) sandwich type 1, (c) sandwich type 2, (d) $\gamma$ and (e) $\beta$. The numbers in the parenthesis indicate the number of similar structures identified in the 2007 data set.}
\end{figure}

\subsubsection{Other common prototypes}
We also checked other common prototypes. The results are summarized in Fig. \ref{Fig8}. 

\textbf{BENZEN}. The crystal structure of benzene is an outlier of herringbone and was less investigated in terms of packing. It was considered to have a pseudo-fcc arrangement due to strong X-ray reflections at the (111) plane \cite{benzene-rmp-1958}. However, from the projection image as shown in Fig. \ref{Fig8}a, such a type can also be explained to have four strongly interacted molecules surrounding the center molecule forming the equator plane, while four secondary interactions are evenly distributed in each of the Northern and Southern Hemispheres. In our explanation, we take into account the fact that the 12 neighboring molecules do not have the same interactions with the centering molecules, which is different from a typical fcc atomic crystal. To our knowledge, there are no previous reports on the crystal sharing the same packing pattern with BENZEN. Here our analysis suggests that at least 14 crystals belong to this family with the threshold distance value of 0.89, such as NUNCUW (0.966) and FAPZOL (0.962). 

\textbf{Sandwich}. As shown in Figs. \ref{Fig8}b-c, both PYRENE and PERlEN01 has one very bright spot in the upper left of the projection image, however, other less bright spots notably differ from each other. Such subtle differences can be easily ignored when one just focuses on the common projection plane. The $S$ between PYRENE and PERLEN01 is only 0.753. Among the entire dataset, such packings do not appear often, with only 3 PYRENE and 4 PERLEN03 types. 

$\bm{\gamma/\beta}$. Two representative crystals (CORONE and TBZPYR), as shown in Figs. \ref{Fig8}d-e, are nearly identical in terms of the strong characteristics. Both CORONE (210) and TBZPYR (186) can find a lot of structures sharing the same packing pattern. The $S$ value between CORONE and TBZPYR is 0.975. Therefore, these two families are expected to have a lot of overlaps. %Again, one can use the fine resolution model to search for really matched crystals.

\begin{figure}[ht]
\centering
\includegraphics[width=0.5\textwidth]{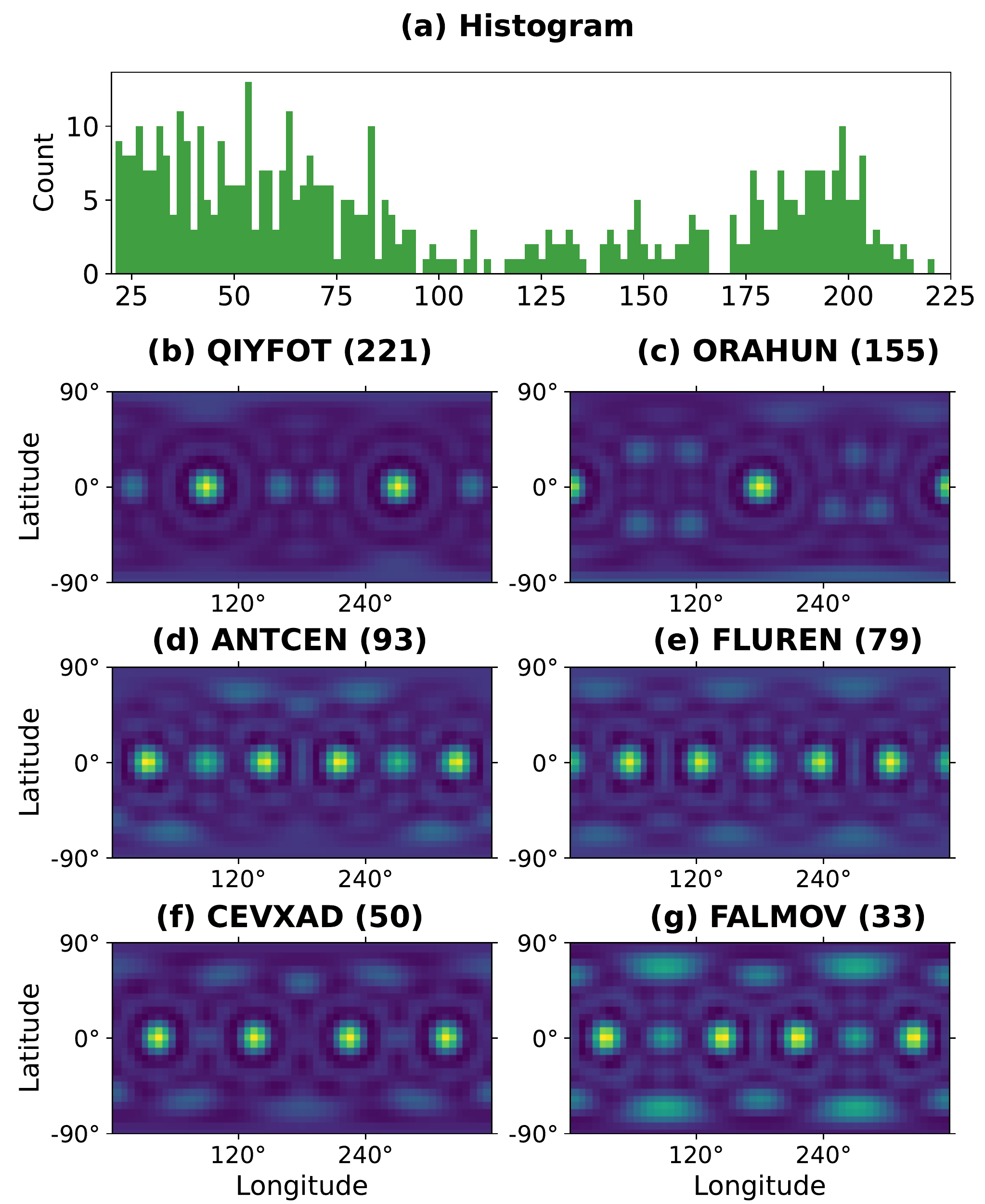}
\caption{\label{Fig9} The complete clustering results and several abundant packing motifs. (a) is the histogram of similar structures for each crystal with a threshold of 0.88. (b)-(h) show six representative structures' patterns featured by two, four, and six bright spots in the projected equatorial plane. The numbers in the parenthesis indicate the number of similar structures identified in the 2007 data set.}
\end{figure}

\subsubsection{A complete categorization}
\begin{figure}[ht]
\centering
\includegraphics[width=0.45\textwidth]{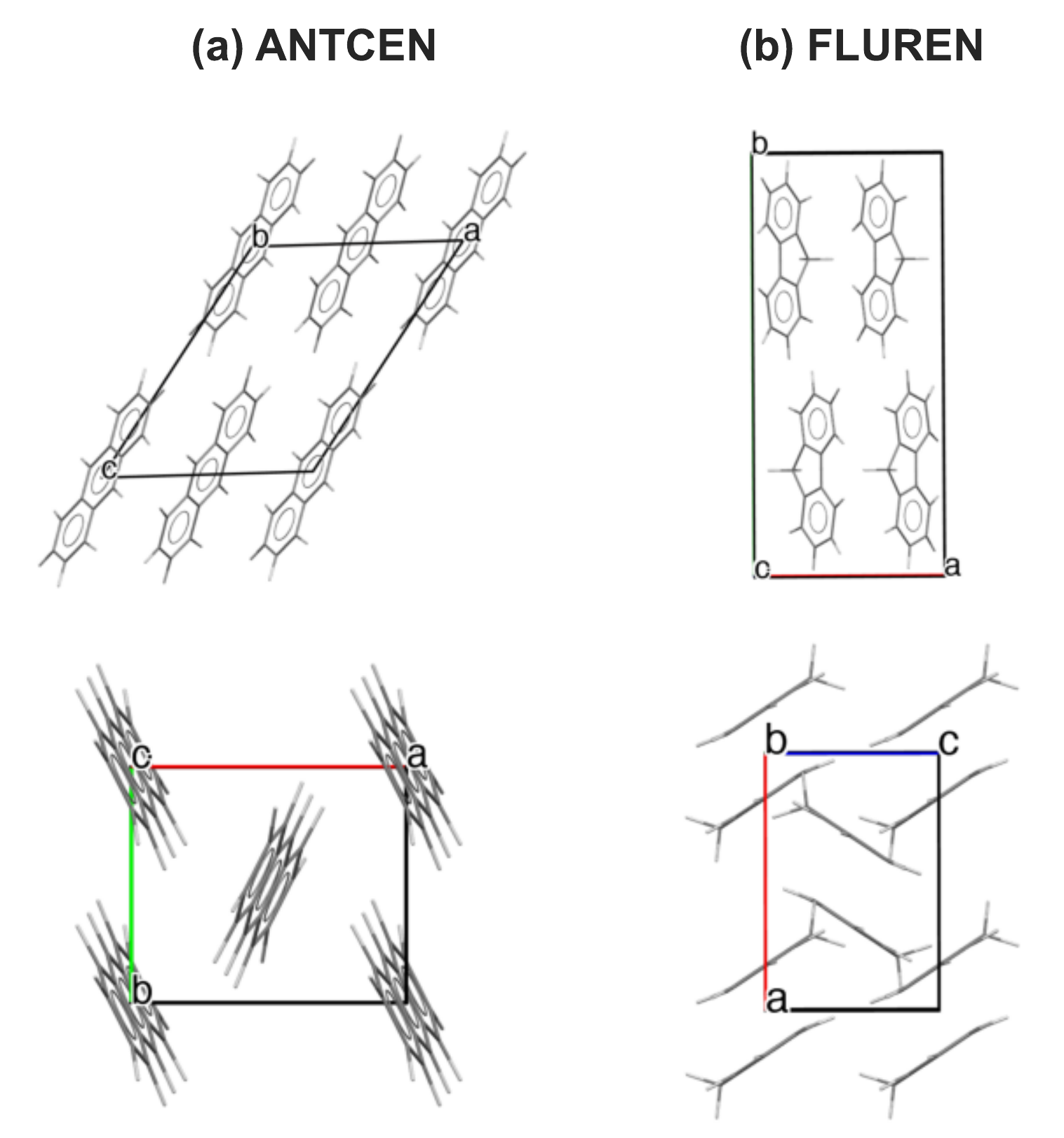}
\caption{\label{Fig10} Crystal packing comparison between (a) ANTCEN and (b) FLUREN.}
\end{figure}

We also performed a systematic pairwise distance analysis for the entire 2007 crystal data set. Using the threshold of 0.88, we count the number of similar structures for each crystal as shown in  Fig. \ref{Fig9}a. From the distribution, it is clear that many structures possessing the pattern of two strong bright spots (see Figs. \ref{Fig9}b-c) are most abundant. This pattern is also consistent with what we have observed in $\gamma/\beta$. However, we note that other neighbors in the crystal can adopt different packing sequences. Therefore, there exist many subtypes within this family. The second large group is featured by the pattern of six bright spots forming the equatorial plane (see Figs. \ref{Fig9}d-e)). Therefore, the most important feature of this group is that the molecules form close packing layers with strong intermolecular interactions \cite{Kitaigorodskii-JAC-2003}. Within this group, ANTCEN belongs to the common naphthalene family. However, the projection map in Fig. \ref{Fig9}d is rotated to put the six bright spots in the equatorial plane. Another frequent-occurring subgroup is shown in Fig. \ref{Fig9}e of FLUREN. The structure differs from ANTCEN by the relative shifting of molecules in the adjacent layers. As shown in Fig. \ref{Fig10}, the strong similarity between ANTCEN and FLUREN can be easily detected in real space. However, it can also be seen that ANTCEN follows the herringbone pattern and FLUREN adopts the sandwich type from other choices of projection. This drastic difference clearly suggests the limitation of interpreting structure by visual projection. On the other hand, our spherical image representation provides more rigorous comparison by capturing both similarity and dissimilarity at all directions. Finally, we also observe another group of structures with four bright spots in the equatorial plane (see Figs. \ref{Fig9}g-h). This is less common. Each structure of this group has only 30-50 similar structures within the 2007 data set. 

\section{Conclusions}
In sum, we present an image approach to characterize and classify molecular packing from the local neighboring environment. The similarity between two crystals (i.e., images) can be evaluated from the spherical harmonics expansion based on the maximum cross-correlation. We apply this approach to investigate the packing similarity analysis over 2000 C-H crystal data sets and categorize them by similarity. Compared to the previous approaches based on visual comparison in the real space, our spherical harmonics expansion provides a robust way to measure the packing similarity. It is suitable for a rapid search for similar crystal structures by crystal packing from a large data set. For instance, there is a general perception that herringbone is high-mobility \cite{Mattheus2001,Jurchescu2006,Izawa2008}. Our tool can clearly be applied to search for new candidates with a similar packing pattern. In the future, a more extensive crystal packing motif library can be built by using this criterion. When such templates becomes available, it can map the relation between molecule properties and crystal packing to guide the design of new materials. Alternatively, it can also be used to complement the crystal structure prediction by chemical substitution of the well defined structural prototypes \cite{Reilly-Acta-2016}.

\section*{Acknowledgments}
Q.Z. acknowledge the NSF (DMR-2142570) and Sony Group Corporation for their financial supports. The computing resources are provided by XSEDE (TG-DMR180040).

\appendix

\section{Real Spherical Harmonics}
The real spherical harmonics are defined as

\begin{equation}
    Y_{lm} (\theta, \phi) = 
    \begin{cases}
    \Bar{P}_{lm}(\cos{\theta}) \cos m\phi,       & \text{if } m\geq 0\\
    \Bar{P}_{l|m|}(\cos{\theta}) \sin |m|\phi,   & \text{if } m < 0
\end{cases}
\end{equation}

where normalized associated Legendre functions with the 4$\pi$-normalized spherical harmonic functions are given by
\begin{equation}
    \Bar{P}_{lm}(\mu) = \sqrt{(2-\delta_{m0}(2l+1) \frac{(l-m)!}{(l+m)!})} P_{lm}(\mu)
\end{equation}

and $\delta_{ij}$ is the Kronecker delta function. 

The unnormalized associated Lengendre functions are from the following relations
\begin{equation}
\begin{split}
    P_{lm}(\mu) &= (1-\mu^2)^{m/2} \frac{d^m}{d\mu^m} P_l(\mu)\\
    P_l(\mu) &= \frac{1}{2^l l!} \frac{d^l}{d\mu^l} (\mu^2 - 1)^l    
\end{split}
\end{equation}

In this work, we strictly follow the definitions used in the \texttt{pyshtools} package \cite{shtools}. The image conversion and optimization functions have been implemented in the \texttt{PyXtal} package \cite{pyxtal}.

\section*{Contributions}
O.Z. conceived the idea, wrote the code, and performed the simulation; W.T. participated in code development, Q.Z and H.S wrote the paper.

\section*{Data availability}
The data sets generated during and/or analysed during the current study are available from the corresponding author on reasonable request. 

\section*{Code availability}
The codes used to calculate the results of this study are available in \url{https://github.com/qzhu2017/PyXtal}.

\section*{Conflict of interest}
Q.Z. have received research funding from Sony Group Corporation.
%All authors declare that they have no conflict of interest.

\nolinenumbers
\section*{REFERENCES}
\bibliography{ref}
\end{document}